\documentclass[10pt, conference]{IEEEtran}
\usepackage{blindtext}
\usepackage{graphicx}
\usepackage{hyperref}
\usepackage{amssymb}
\usepackage{amsfonts}
\usepackage{multicol}
\usepackage{amsmath}
\usepackage{mathtools}
\usepackage[mathscr]{euscript}
\hypersetup{colorlinks=true, citecolor=red}
\usepackage{hyperref}
\usepackage{booktabs}
\usepackage{makecell}
\usepackage{multirow}
\usepackage{amsbsy} 
\usepackage[hypcap=true, font=small]{caption}
\usepackage[lined,vlined,ruled,linesnumbered]{algorithm2e}

\hypersetup{
    colorlinks=true,
    linkcolor=red,
    filecolor=black,      
    urlcolor=red,
}

\DeclareMathOperator*{\argmin}{argmin}
\newcommand*{\argminl}{\argmin\limits}

\DeclareMathOperator*{\argminrr}{\Theta}
\newcommand*{\argminr}{\argminrr\limits}

\usepackage{color, colortbl}
\definecolor{Gray}{gray}{0.9}
\usepackage[first=0,last=9]{lcg}

\usepackage{tikz}

\newcommand\copyrighttext{%
  \footnotesize \textcopyright~\the\year~IEEE. Personal use of this material is permitted.
  Permission from IEEE must be obtained for all other uses, in any current or future
  media, including reprinting/republishing this material for advertising or promotional
  purposes, creating new collective works, for resale or redistribution to servers or
  lists, or reuse of any copyrighted component of this work in other works.}
\newcommand\copyrightnotice{%
\begin{tikzpicture}[remember picture,overlay]
\node[anchor=north,yshift=-15pt] at (current page.north) {\fbox{\parbox{\dimexpr\textwidth-\fboxsep-\fboxrule\relax}{\copyrighttext}}};
\end{tikzpicture}
\vspace{-0.3cm}
}

\begin{document}

\title{Distributed VNF Scaling in Large-scale Datacenters: 
An ADMM-based Approach\vspace{-2.5ex}}

\author{\IEEEauthorblockN{Farzad Tashtarian$^\star$, Amir Varasteh$^\dagger$, Ahmadreza Montazerolghaem$^\S$, and Wolfgang Kellerer$^\dagger$} \\ 
$^\star$ Department of Computer Engineering, Mashhad Branch, Islamic Azad University, Mashhad, Iran\\ \{Email: f.tashtarian@mshdiau.ac.ir\} \\
$^\dagger$ Chair of Communication Networks, ِDepartment of Electrical and Computer Engineering, \\Technical University of Munich, Munich, Germany\\ \{Email: amir.varasteh, wolfgang.kellerer@tum.de\}\\
$^\S$ Department of Computer Engineering, Ferdowsi University, Mashhad, Iran \\ \{Email: ahmadreza.montazerolghaem@stu.um.ac.ir\}}

\maketitle
\copyrightnotice{}

\begin{abstract}

Network Functions Virtualization (NFV) is a promising network architecture where network functions are virtualized and decoupled from proprietary hardware. In modern datacenters, user network traffic requires a set of Virtual Network Functions (VNFs) as a service chain to process traffic demands. Traffic fluctuations in Large-scale DataCenters (LDCs) could result in overload and underload phenomena in service chains. In this paper, we propose a distributed approach based on Alternating Direction Method of Multipliers (ADMM) to jointly load balance the traffic and horizontally scale up and down VNFs in LDCs with minimum deployment and forwarding costs. Initially we formulate the targeted optimization problem as a Mixed Integer Linear Programming (MILP) model, which is NP-complete. Secondly, we relax it into two Linear Programming (LP) models to cope with over and underloaded service chains. In the case of small or medium size datacenters, LP models could be run in a central fashion with a low time complexity. However, in LDCs, increasing the number of LP variables results in additional time consumption in the central algorithm. To mitigate this, our study proposes a distributed approach based on ADMM. The effectiveness of the proposed mechanism is validated in different scenarios.
\end{abstract}

\begin{IEEEkeywords}
Network function virtualization, datacenters, VNF scaling, service chaining, distributed optimization, alternating direction method of multipliers (ADMM).
\end{IEEEkeywords}

\IEEEpeerreviewmaketitle

\section{Introduction}

\begin{figure}[tp]
\includegraphics[width=\linewidth]{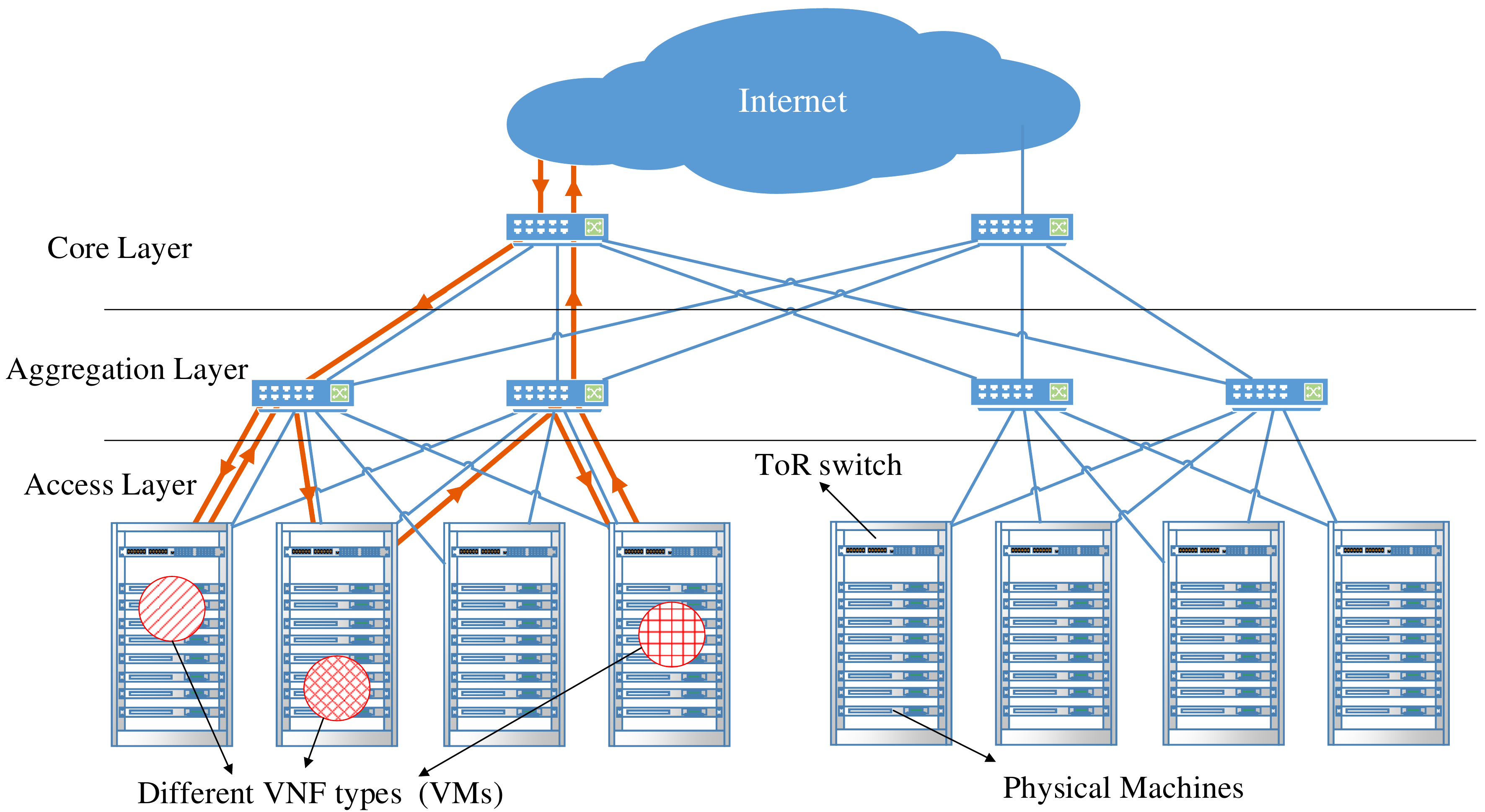}
\caption{A service chain with three VNFs in a conventional datacenter}
\label{fig1}
\end{figure}

Today, different technologies such as Software Defined Networking (SDN) and Network Function Virtualization (NFV) are being integrated to facilitate the modern datacenter service providers \cite{mijumbi2015network}. In these datacenters, the network services are frequently deployed as Virtual Network Function (VNF) chains to process the incoming traffic \cite{joseph2008policy}  (see Fig. \ref{fig1}). The NFV architecture relies on off-the-shelf Physical Machines (PMs) which provide computational resources to several Virtual Machines (VMs), each VM implementing a network function with special software programs. In addition, SDN technology could manage the communication network between these VNFs in the datacenter. \par

Due to traffic fluctuations in datacenters, resource over and underload can occur in service chains. The impact of an overload could be the degradation of overall performance and SLA violations, while underload could lead to significant resource wastage. One of the most popular methods to ensure efficient resource management, is the appropriate placement of VNFs on PMs, which is called the VNF placement problem \cite{basta2014applying}. This problem has been addressed in literature regarding various parameters, such as network delay and end-to-end latency \cite{krishnaswamy2015latency,thai2016joint}, and operational costs of VNFs (deployment cost, forwarding/processing cost, and energy consumption) \cite{bari2015orchestrating, cohen2015near}, and \cite{ghaznavi2015elastic}. Although the majority of studies focus on the initial placement of the service chain, network-aware resource management of the service chains during their operational time has emerged as a critical issue. To overcome this issue, the traffic could be load balanced between VNF instances, or the over and underloaded VNF(s) could be scaled up and down. In this paper, we jointly load balance the traffic and horizontally scale up and down VNFs with minimum deployment and forwarding costs. Notably, we use thresholds to automatically detect over and underloaded VM(s), respectively.\par

Based on the above considerations, the problem is modeled using Mixed Integer Linear Programming (MILP) which is NP-complete \cite{vavasis1991nonlinear}. Therefore, we relax our MILP model into two LP formulations. These models perform well in small and medium size datacenters. However, in Large-scale DataCenters (LDCs), the number of variables and constraints of the LP models escalate significantly resulting in an increase in the time complexity of the algorithm, which requires the use of a distributed approach. Hence, we extend our approach to distributed optimization, using a technique known as Alternating Direction Method of Multipliers (ADMM) \cite{boyd2011distributed}. In addition to simplicity and its powerfulness, this method has several advantages like fast convergence, low computational complexity, and low overhead of message passing \cite{boyd2011distributed}. Using this method, the convex optimization problem is decomposed into different subproblems, each of which is easier to handle. These subproblems are solved iteratively and in parallel using defined agents \cite{boyd2011distributed}. 
\par Although the conventional ADMM is guaranteed to converge for two-block optimization problems, for \textit{n}-block variable problems (i.e. multi-block variables) with $n>2$, updating different variables sequentially has no convergence guarantee \cite{chen2016direct}. Therefore, in this paper, because our problem is in form a of multi-block ADMM, we use a specific ADMM extension called Randomly Permuted ADMM (RP-ADMM) \cite{sun2015expected} to solve the problem in a distributed way.

\par Afterall, considering the aforementioned explanations, the main contributions of our proposed work are described as follows:
\begin{enumerate}
  \item Introducing a joint traffic load balancing and VNF scaling mechanism to mitigate both over and underloaded situations in datacenters,
  \item Formulating the above problem as a MILP optimization model, and proposing two LP-relaxations to address time complexity,
  \item Extending our approach to a distributed method based on the RP-ADMM technique,
  \item Evaluation of the proposed work in different scenarios via simulation.
\end{enumerate} \par

The remainder of this paper is organized as follows, we start by presenting the related work in the literature (section II). Then, in sections III and IV we illustrate the system model and problem formulation, respectively. Next, the proposed approaches are described in section V. We validate our approach in section VI, and finally, section VII concludes the whole paper.

\section{Related Work}

VM placement has been widely studied in the cloud computing environment \cite{jiang2012joint, meng2010improving, varasteh2015server}, and \cite{basta2014applying}. However, specific characteristics and constraints of the NFV paradigm such as service chaining requirements, deployment and license costs, etc., has made the VNF placement problem more complex. Several studies exist in the literature that tackle this problem. For instance, StEERING \cite{zhang2013steering} focused on VNF chain placement and traffic steering and uses a simple heuristic algorithm to solve the problem. Bari \textit{et. al.} \cite{bari2015orchestrating} proposed a model that can be used to determine the optimal number of VNFs and to place them at optimal locations (PMs). In \cite{cohen2015near}, authors modeled this problem to minimize client-server communication distance and operational costs. Also, \cite{luizelli2015piecing} presented a joint traffic steering and VNF placement using an Integer Linear Programming (ILP) model which sought to minimize the total delay of the traffic. While these approaches largely focus on initial placement of VNFs, they do not monitor VNF chains during their operational time. Hence, because of traffic fluctuations, a resource over or underload could happen in VNF chains, which is a significant challenge to address. \par

The work most similar to our approach is \textit{Stratos} \cite{gember2013stratos}. It uses a four-step solution attempting to solve the VNF chain bottlenecks during its operational time. When an overload happens, it initially uses a flow distribution method. When this method fails to solve the bottleneck, \textit{Stratos} attempts to migrate the VNFs. If the problem persists, \textit{Stratos} scales up the bottlenecked VNFs. Finally, when all these procedures failed, it scales up all the VNFs in the chain with a fixed number of instances. The primary drawback of \textit{Stratos} is that it works in a trial manner, which creates resource wastage and inefficiencies to solve the resource bottlenecks. Another close work to us is \cite{wang2016online} which attempts to tackle dynamic provisioning of one or multiple service chains in cloud datacenters in a centralized and time-slotted manner. However, \cite{wang2016online} does not address the forwarding cost in their model. This could bring additional growth of east-west traffic in the datacenter which increases the wastage of datacenter computation resources \cite{chi2015efficient}.\par

In contrast to afore-mentioned works, we introduce a joint traffic load balancing and VNF scaling mechanism to deal with over and underload phenomena during service chain operational time, while effectively considering forwarding and deployment costs in form of a mathematical optimization model. To the best of our knowledge, there is no literature that is applicable in LDCs. Therefore, we extend the model to a distributed form of optimization, using RP-ADMM method, to be able to use this approach in LDCs.

\begin{figure*}[t]
\centering
\includegraphics[width=\textwidth]{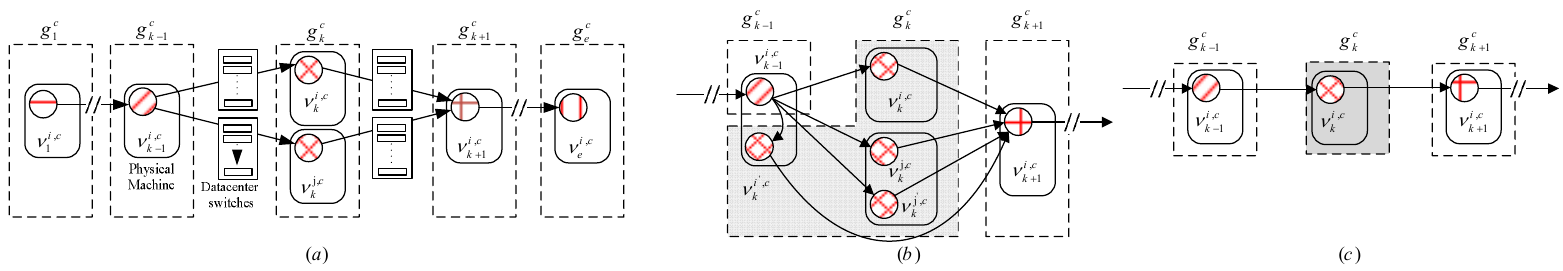}
\caption{VNF service chain states in a datacenter network, (a) Normal state: an operational VNF chain in a network; (b) Overload occurrence: VMs $i'$ and $j'$ are launched to manage the overload situation; and (c) Underload occurrence: VM $j$ is turned off to optimize resource usage efficiency in the VNF chain.}
\label{fig2}
\end{figure*}

\section{System Model}

We consider a datacenter as a graph $\mathscr{G}=\{\mathbb{V},\mathbb{L}\}$, where the vertices set $V$ consists of $U$ physical machines $\mathscr{P}_i$ ,$i=1\ldots U$ and $S$ switches $\mathscr{S}_i$ ,$i=1\ldots S$, and $\mathbb{L}$ represents edges of $\mathscr{G}$ where $l_{ij}=1$ if a direct communication link exists between two entities $i,j\in\mathbb{V}$. For the ease of problem formulation, we assume a conventional network topology for the datacenter which consists of three layers: access layer, aggregation layer, and core layer (see Fig. \ref{fig1}). The total available resources of a PM is considered as $\mathscr{P}_i^r$ where $r\in R$ and $R$ is the set of limited types of PM resources such as CPU, memory, bandwidth, etc. Let $C$ be a set of service chains that each of them is constituted by an ordered sequence of VNFs $f_k$, where $k\in K$ and $K$ is the set of various network functions such as Firewall, IDS, NAT, etc. \par

We assume that each VM is providing only one network function. Also, we assume all configured VMs providing network function $k$, are using an identical computational resource. Each VM $i$ that belongs to chain $c\in C$ and provides $f_k$, denoted by $v_k^{i,c}$, is configured with initial resource $u_{k,r}^{i,c}$ ($r\in R$). \par

We use the notation $g_k^c = \{v_j^{i,z}\mid \forall j=k,z=c\}$ indicating a set of VMs that provide $f_k$ on chain $c\in C$. In addition, \textit{ingress} and \textit{egress} points of $g_k^c$  are defined as $g_{k-1}^c$  and $g_{k+1}^c$, where $k+1$ and $k-1$ refer to the \textit{previous} and \textit{next} service function of $f_k$ in chain $c$, respectively. We consider two main phases for each chain: \textit{setup} phase and \textit{operational} phase. In setup phase, the requested service chain is deployed in datacenter network with respect to SLA and the available resources of the datacenter. In the second phase, the chain begins to process the incoming traffic, which rate may vary from time to time. Therefore, by having an initial VNF placement for service chains, in this paper, we focus on the second phase of chain lifetime and address the problem of managing the over and underloaded chain with the minimum deployment and forwarding costs.

\par Our proposed system architecture includes four main components: Monitoring, SDN Controller, VNF Placement and Scaling Module, and REsource Management (REM). Since the proposed approach should respond efficiently and quickly to traffic fluctuations, we use monitoring component to continuously collect the amount of CPU, memory, and bandwidth usage of VMs. In addition to the reported data by monitoring component, the SDN controller collaborates with REM through sharing the routing paths and configuration of network switches. It also updates and configures the routing policy using the REM output. We define $V$ and $V^*$ as the set of running VMs in $g_k^c$ (\textit{online VMs}), and the set of configured VMs to join $g_k^c$ (\textit{offline VM pool}), respectively. Also, let $\tilde{u}_{k,r}^{i,c}$ be the utilization of resource type $r$ of $v_k^{i,c}$ in $\tau$ units of time that is measured by monitoring module. Further, we define $\hat{u}_{k,r}^{i,c}$, $\bar{u}_{k,r}^{i,c}$, and $\check{u}_{k,r}^{i,c}$ as \textit{hot}, \textit{warm}, and \textit{cold} thresholds on resource type $r$ for $v_k^{i,c}$ to enable us to specify the state of a chain, respectively, as follows:\\
\noindent (1) \textit{overload} state: 
\begin{align}
    \exists k : g_k^c \in c \mid \exists i : v_k^{i,c} \in g_k^c \text{ and } \tilde{u}_{k,r}^{i,c} \geq \hat{u}_{k,r}^{i,c} \nonumber
\end{align}
\noindent (2) \textit{underload} state:
\begin{align}
    \exists k : g_k^c \in c \mid \frac{1}{ \mid V \mid } \sum_{i:v_k^{i,c} \in g_k^c} \nolimits \tilde{u}_{k,r}^{i,c} \leq \check{u}_{k,r}^{i,c} \nonumber \\
    \text{and } max\{ \tilde{u}_{k,r}^{i,c} \mid \forall i:v_k^{i,c} \in g_k^c \} \leq \bar{u}_{k,r}^{i,c}  \nonumber
\end{align}
\noindent (3) \textit{normal} state: If chain $c$ is neither in overload nor underload state, it is considered being in the \textit{normal} state. \\
\noindent For illustration, suppose chain $c$ consists of three functions: $f_1$, $f_2$, and $f_3$, where $v_1^{1,c}$, $\{v_2^{1,c},v_2^{2,c}\}$, and $v_3^{1,c}$ are serving these functions, respectively. We assume $50\%$, $65\%$ as CPU utilization of $v_1^{1,c}$ and $v_3^{1,c}$, respectively. In this example, we focus on $f_2$ to show how the state of the chain $c$ can be determined. Therefore, we consider the following two scenarios for different CPU utilization of the VMs serving $f_2$. Also, we assume \textit{hot}, \textit{warm}, and \textit{cold} thresholds as $90\%$, $80\%$ and $30\%$, respectively. Scenario (a) $\{\tilde{u}_{2,[cpu]}^{1,c}=95\%,\tilde{u}_{2,[cpu]}^{2,c}=75\%\}$: Since the CPU utilization of $v_2^{1,c}$ violates the \textit{hot} threshold, chain $c$ falls into the \textit{overload} state. We note that the presence of only one violated VM is sufficient to put a chain in the \textit{overload} state. Scenario (b) $\{\tilde{u}_{2,[cpu]}^{1,c}=40\%,\tilde{u}_{2,[cpu]}^{2,c}=15\%\}$: In this scenario, it can be seen that the average of CPU utilizations is less than the cold threshold. Consequently, this scenario makes chain $c$ to be considered as \textit{underload}. Notably, $f_1$ and $f_3$ cannot be in \textit{underload} state, since there is only one VM assigned for each of these functions.

\par Remarkably, the values of \textit{hot}, \textit{warm} thresholds should be opted in such a way that the gap between their values provides a safety margin to cope with the temporary network traffic fluctuations. Additionally, different types of resources can have different thresholds. For example, the hot thresholds for CPU and memory resources can be set to $90\%$ and $80\%$, respectively \cite{xiao2013dynamic}. In Fig. \ref{fig2}, we illustrate the three operational states for a sample chain.

\section{Problem Formulation}

In this section, we address the problem of overload and underload events that occur in network service chain. First, we present a central optimization model, and then extend our study by proposing a distributed model using the ADMM technique. Table \ref{tab:table1} shows the main notations.\par

To deal with the over/underloaded chain $c$, a number of constraints must be satisfied. Suppose $g_k^c$ is identified as an over/underloaded set in chain $c$. Let $b_{ij}$ be a binary decision variable, where if $b_{ij}=1$, then VM $j\in \{V\cup V^*\}$ must be running on $\mathscr{P}_i$. We denote $\alpha_{ij}$ as the percentage of collaboration of VM $j\in \{V\cup V^*\}$ on $\mathscr{P}_i$, with existing VMs $\in V$ to process the total traffic coming from $g_{k-1}^c$. The value of $\alpha_{ij}$ must be restricted as follows:
\begin{align}
   &\varepsilon b_{ij}\leq\alpha_{ij}\leq \varphi_k b_{ij}, \qquad \forall j\in \{V\cup V^*\}, \forall i\in \mathscr{P}  \label{eq1}
\end{align}
where $\varepsilon$ shows the minimum amount of collaboration that causes a VM to be launched on a PM. Additionally, $\varphi_k$ shows the maximum percentage of service capacity of VMs running network function $k$ in such a way that $\forall j\in \{V\cup V^*\}$, $\forall r\in R$, $\varphi_k \omega_k^r T<u_{k,r}^{j,c}$, where $\omega_k^r$ is defined as the service impact factor of network function $k$ on resource $r\in R$. Let $n_{ij}^d$  and $m_{ij}^d$, $(i,j\in \{\mathscr{S}\cup \mathscr{P}\}$  and $d\in \{V \cup V^*\})$, be the amount of unprocessed traffic rates destined to $v_k^{d,c}$, and processed traffic by $v_k^{d,c}$ transmitting from $i$ to $j$, respectively. The following constraints state that all online VMs must cover the total traffic flow originated by ingress point $g_{k-1}^c$:
\begin{align}
   &\sum_{j\in \{V\cup V^*\}} \nolimits \sum_{i\in \mathscr{P}} \nolimits \alpha_{ij}=1 \label{eq2} \\ 
   \sum_{s\in \{\mathscr{S}\cup \mathscr{P}\}} \nolimits &l_{si}n_{si}^j=T\alpha_{ij}, \qquad \forall j\in \{V\cup V^*\}, \forall i\in \mathscr{P} \label{eq3}
\end{align}
\noindent
where $T$ is the total traffic coming from $g_{k-1}^c$ (i.e. \textit{ingress} point) to the over/underloaded $g_k^c$. The following constraint indicates that the total required resources for deploying VM(s) on $\mathscr{P}_i$ must not be more than its available resources, $\mathscr{P}_i^r$:
\begin{align}
    \sum_{j\in \{V\cup V^*\}} \nolimits b_{ij}u_{k,r}^{j,c}\leq \mathscr{P}_i^r, \hspace{1cm} \forall i\in \mathscr{P}, r\in R \label{eq4}
\end{align}
Additionally, $\forall i\in \mathscr{S} \text{ and } \forall d\in \{V\cup V^*\},$ the next constraint is presented as follows: 
\begin{align}
   &\sum_{j\in\{\mathscr{S}\cup \mathscr{P}\}} \nolimits l_{ij}(n_{ij}^d-n_{ji}^d)=0 \label{eq5}\\
   &\sum_{j\in\{\mathscr{S}\cup \mathscr{P}\}} \nolimits l_{ij}(m_{ij}^d-m_{ji}^d)=0 \label{eq6}
\end{align}
These constraints state that the total incoming (processed and unprocessed) traffic to $\mathscr{S}_i$ must be equal to the total outgoing traffic from it. A similar constraint $\forall i\in \mathscr{P}$ and $\forall d\in  \{V\cup V^*\}$ could be written as follows:
\begin{equation}
    \sum_{j\in \{\mathscr{S}\cup \mathscr{P}\}} \nolimits l_{ij} ( m_{ij}^d - \gamma_k n_{ji}^d  ) =0  \label{eq7}
\end{equation}
where $\gamma_k >0$ is data changing factor of $f_k$ serviced by $g_k^c$ \cite{ma2015traffic}. Although a connection between any pairs of PMs in the datacenter must be established via a switch, the data flow between two different types of VMs, both hosted by an identical PM, is applicable. The following constraints guarantee that the total traffic $T \gamma_k $, processed by $g_k^c$, is received by $g_{k+1}^c$ (i.e. \textit{egress} point):
\begin{align}
   &\sum_{d\in \{V\cup V^*\}} \nolimits \sum_{i:\mathscr{P}_i\in g_{k-1}^c} \nolimits \sum_{j\in \{\mathscr{S}\cup \mathscr{P}\}} \nolimits l_{ij}n_{ij}^d = T \label{eq8} \\
   &\sum_{d\in \{V\cup V^*\}} \nolimits \sum_{i:\mathscr{P}_i\in g_{k+1}^c} \nolimits \sum_{j\in \{\mathscr{S}\cup \mathscr{P}\}} \nolimits l_{ji}m_{ji}^d = T\gamma_k \label{eq9}
\end{align}
Moreover, $\forall i,j \in \{\mathscr{S} \cup \mathscr{P}\}$ a bandwidth constraint could be formulated as follows:
\begin{align}
    \sum_{d:v_k^{d,c}\in g_k^c} \nolimits l_{ij}(n_{ij}^d+m_{ij}^d) \leq w_{ij} \label{eq10}
\end{align}
where $w_{ij}$ is the total available bandwidth of link$(i,j)$. To evaluate the performance of any solutions that satisfies the above constraints, we now propose the following two cost functions:
\begin{align}
   &\text{$Deployment \text{ } Cost:$ } \nonumber \\
   &\sum_{i\in \mathscr{P}} \nolimits \sum_{j\in V^*} \nolimits b_{ij}\mathbb{X} - \sum_{i\in \mathscr{P}} \nolimits \sum_{j\in V} \nolimits b_{ij}\hat{\mathbb{X}} \label{eq11}\\
    &\text{$Forwarding \text{ } Cost:$ } \nonumber \\
   &\sum_{d\in \{V\cup V^*\}} \nolimits  \sum_{i\in \{\mathscr{S}\cup \mathscr{P}\}} \nolimits \sum_{j\in \{\mathscr{S}\cup \mathscr{P}\}} \nolimits  F(i,j) (n_{ij}^d+m_{ij}^d) \label{eq12}
\end{align}

Also, $\mathbb{X}$ and $\hat{\mathbb{X}}$ are considered as two penalty values in the deployment cost formulation. We use the deployment cost to control the power states, migration, and placement of VMs. In fact, it can be utilized for both over and underload situations by adjusting the proper values $\mathbb{X}$ and $\hat{\mathbb{X}}$ (see Proposition. 1 for proof). Further, we use the forwarding cost to force the model to deliver traffic via efficient routes. To achieve this, we should prevent the traffic from being transferred through higher layers of network in the topology (i.e. aggregation and core layers, see Fig. \ref{fig1}). Hence, we defined $F(i,j)$ as the forwarding cost which applies on transmitting composite bit-stream per unit of time, and it is proportional to layers of datacenter topology in which $i$ and $j$ are located. Finally, the following MILP is proposed to jointly balance the traffic load and scale VNFs with the minimum costs:
\begin{align}
    &\textit{\textbf{minimize}} \qquad \aleph_1 Eq. (\ref{eq11})+ \aleph_2 Eq. (\ref{eq12}) \label{eq13}\\
    &\textit{\textbf{s.t.}} \nonumber \\
    &Constraints \text{  } (\ref{eq1}) - (\ref{eq10}) \nonumber\\
    &\textit{\textbf{vars. }}b_{ij}\in \{0,1\},\alpha_{ij},n_{ij}^d,m_{ij}^d\geq 0 \nonumber
\end{align}
\noindent
where $\aleph_1$ and $\aleph_2$ are two constant weights for deployment and forwarding costs.\par

\begin{table}[t]
    \centering
    \small
    \caption {Notations}
    \label{tab:table1}
    \begin{tabular*}{1\linewidth}{cl}
    \Xhline{2\arrayrulewidth}
    Notation & \hspace{2.5cm} Description \\ \hline
    $\mathscr{S}_i$ & The $i^{th}$ switch in datacenter network \\
    $\mathscr{P}_i$ & The $i^{th}$ physical machine in datacenter network \\
    $R$ & The set of resource types (CPU, Memory, etc.) \\
    $K$ & The set of network function types \\
    $f_k$ & The network function type $k\in K$ \\
    $l_{ij}$ & $l_{ij}=1$ if there is a link between two entities \\
    & $i,j\in \{\mathscr{S} \cup \mathscr{P}\}$\\
    $\mathscr{P}_i^r$ & The total available amount of resource  $r\in R$ in $\mathscr{P}_i$ \\
    $v_k^{i,c}$ & VM $i$ that belongs to chain $c$ and hosts $f_k$ \\
    $g_k^c$ & The set of VMs that provide $f_k$ on chain $c$ \\
    $g_{k-1}^c$, $g_{k+1}^c$ & The ingress/egress points of $g_k^c$ \\
    $\tilde{u}_{k,r}^{i,c}$ & The utilization of $r^{th}$ resource of $v_k^{i,c}$ \\
    $\hat{u}_{k,r}^{i,c}$, $\bar{u}_{k,r}^{i,c}$, $\check{u}_{k,r}^{i,c}$& \textit{Hot}, \textit{warm}, and \textit{cold} thresholds for $r^{th}$ resource of $v_k^{i,c}$ \\
    $u_{k,r}^{i,c}$ & The initial resource type $r$ for $v_k^{i,c}$ \\
    $n_{ij}^d$ & The amount of unprocessed traffic destined to $v_k^{d,c}$ \\ ~ & transmitting from $i$ to $j$, where $i,j\in \{\mathscr{S}\cup \mathscr{P}\}$ \\
    $m_{ij}^d$ & The amount of processed traffic by $v_k^{d,c}$ transmitting \\
    & from $i$ to $j$ where $i,j\in \{\mathscr{S}\cup \mathscr{P}\}$ \\
    $T$ & The total traffic received by $g_k^c$  \\
    $\gamma_k$ & Traffic changing factor of $f_k$ \\
    $\alpha_{ij}$ & The percentage of collaboration of $v_k^{j,c}$, hosted on \\
    ~ & $\mathscr{P}_i$, with other VMs providing identical network \\
    ~ & function and belong to $g_k^c$  \\
    $\varphi_k$ & The maximum percentage of service capacity of \\ 
    ~ & VMs running network function $k$\\
    $\omega_k^r$ & The service impact factor of network function $k$ on\\
    & resource $r\in R$ \\
    $b_{ij}$ & $b_{ij}=1$ if  VM $j$ is hosted by $\mathscr{P}_i$ \\
    $w_{ij}$ & Total available bandwidth of link$(i,j)$, $i,j\in \{\mathscr{S} \cup \mathscr{P}\}$ \\
    $v^*$ & The number of required VNFs to process the received \\
    & traffic by $g_k^c$ \\
    $V$, $V^*$ & Online VMs belongs to $g_k^c$, offline VM pool for $g_k^c$ \\
    $\Psi$, $\Psi^*$ & The set of PMs hosting $V$ and candidate PMs for\\
    ~& hosting $V^*$ \\
    $e_{ij}$ & The amount of interest of each PM $\in \{ \Psi \cup \Psi^* \}$ in 
    \\ ~& processing the incoming traffic \\
    \Xhline{2\arrayrulewidth}
    \end{tabular*}
\end{table}

\textbf{Proposition 1.} \textit{The proposed MILP model (\ref{eq13}) could be used to mitigate both over and underloaded service chains with adjusting the appropriate penalty values of $\mathbb{X}$ and $\hat{\mathbb{X}}$}.\par
\begin{proof}
Suppose $g_k^c$ is an over/underloaded set which serves VNF type $k$ in chain $c$. In Eq. (\ref{eq11}), we presented the deployment cost that consists of two terms: the first term handles the online VMs, $v_k^{i,c}\in g_k^c$, and the second term focuses on offline VM instances of $k$ which are ready to be transferred to a PM and start servicing in $g_k^c$. In the presence of an overload, there is a solution that jointly balances the traffic rates among the online VMs, and launches one or more offline VMs. Considering overall cost, by choosing two positive values for $\mathbb{X}$ and $\hat{\mathbb{X}}$, where $\hat{\mathbb{X}}\gg \mathbb{X}$, the MILP overcomes overload using flow distribution on online VMs, horizontal scaling, and migration, respectively. On the other hand, when an underload is detected, the best solution is to turn off one or more online VMs and load balance the traffic rates among the remained online VMs. To preserve the resource efficiency, there is no need to turn on new VM instances, unless migrating a VM dramatically decreases the forwarding cost. Hence, the value of $\hat{\mathbb{X}}$ must be set to a large negative number while $\mathbb{X}$ should remain as a very large positive number.
\end{proof}
Considering the aforementioned observations, by detecting over/underloaded chain, the REM adjusts the proper values for $\mathbb{X}$ and $\hat{\mathbb{X}}$ and triggers the optimization algorithm.

\section{The Proposed Approach}

Since the proposed MILP model (\ref{eq13}) is NP-complete and suffers from high time complexity \cite{vavasis1991nonlinear}, we consider two linear relaxations of the MILP model to address over and underloaded VNFs. These LP models could be centrally run in small or even medium size datacenters. However, using these models in LDCs is not applicable due to increasing the number of variables and constraints. Thus, we extend our study by proposing a distributed mechanism based on the RP-ADMM technique.

\subsection{The Central Design}
In case of an overloaded $g_k^c$, suppose that the number of required VNFs $f_k$, which is denoted by $v^*$, to process the received traffic equals to $Max\{\lceil T\omega_k^r/u_{k,r}^{j,c}\rceil \mid \forall r\in R\}$. Also, we define $\Psi$ and $\Psi^*$ sets, which are the set of PMs that are hosting $V$, and the set of candidate PMs that has adequate resources to host at least one VM $\in V^*$, respectively. The relaxed overload model of MILP (\ref{eq13}) is presented as follows:\\ \\
\textbf{\textit{The overload LP model:}}
\begin{align}
 &\textit{\textbf{minimize}} \qquad Eq.(\ref{eq12}) \label{eq14} \\
 &\textit{\textbf{s.t.}} \nonumber \\
 &Constraints \hspace{5.2pt} (\ref{eq5})-(\ref{eq9}) \nonumber \hspace{125pt}\text{(I)-(V)}\\
 &\alpha_{ij}-\alpha_{qj}=0, \hspace{.77cm} i=|V|+1:v^*, \forall j,q\in \{\Psi \cup \Psi^*\} \nonumber \hspace{.2cm}(\text{VI})\\ 
 &\sum_{q\in \{\Psi \cup \Psi^*\}} \nolimits e_{qj}-\alpha_{ij}=0, \ \hspace{.05cm} i=1:v^*, \forall j\in \{\Psi \cup \Psi^*\} \nonumber \text{(VII)}\\ 
 &\sum_{s\in \{\mathscr{S}\cup \mathscr{P}\}} \nolimits l_{si}n_{si}^j-Te_{ij}=0, \nonumber \\
 & \hspace{3.2cm} i=|V|+1:v^*,\forall j\in \{\Psi \cup \Psi^*\} \nonumber \hspace{.1cm} (\text{VIII})\\
 &\sum_{s\in \{\mathscr{S}\cup \mathscr{P}\}} \nolimits l_{si}n_{si}^j-T\alpha_{ij}=0, \hspace{.62cm} i=1:|V|,\forall j\in \Psi \nonumber \hspace{.25cm}(\text{IX}) \\
 &\textit{\textbf{vars. }} \hspace{1cm} n_{ij}^d,m_{ij}^d\geq0,0\leq \alpha_{ij}\leq \varphi_k,0\leq e_{ij}\leq 1 \nonumber
\end{align}

\begin{figure*}[htp]
\centering
\includegraphics[width=\textwidth]{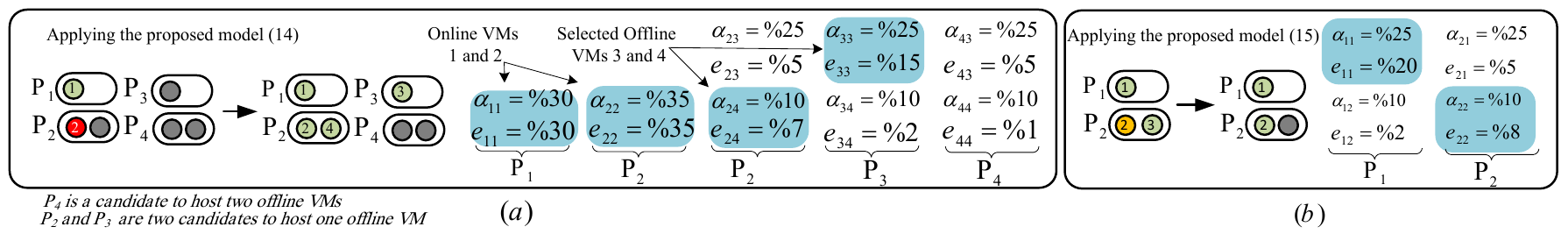}
\caption{An example of over and underloaded chains and their solution, (a) The overload case (b) The underload case}
\label{fig3}
\end{figure*}

Since the number of desired VMs is known, the deployment cost in the objective function could be omitted. In fact, having $v^*$, the relaxed overload model (\ref{eq14}) specifically focuses on the forwarding cost. In this model, the routing constraints could be used as they were before, except the bandwidth constraint (Eq. (\ref{eq10})). We ignored the bandwidth constraint in the proposed relaxed models. However, we will relax this assumption in our future work. Additionally, to relax the binary variable $b_{ij}$ in the MILP model (\ref{eq13}), we define $e_{ij}\in \mathbb{R}^+$ as a new variable in which $j=1:v^*$, and $i\in \{\Psi \cup \Psi^*\}$. This variable identifies the amount of interest of each PM in processing the incoming traffic. For each required instance $j$, the PM with the highest value of $e_{ij}$ is selected to run the $j^{th}$ VM instance and handle $T\alpha_{ij}$ of traffic (constraints (VI)-(VIII)). Finally, constraint (IX) determines the amount of received traffic that must be processed by each PM $i\in \Psi$. \par

On the other hand, when an underload occurs, the following LP-relaxation model, which can be obtained easily from the above model, runs on the first $v^*$ online VMs to appropriately assign them to PMs $\in \Psi$ and distribute received load from $g_{k-1}^c$ on $g_k^c$:\\ \\
\textbf{\textit{The underload LP model:}}
\begin{align}
&\textit{\textbf{minimize}}\qquad Eq.(\ref{eq12}) \label{eq15}\\
&\textit{\textbf{s.t.}} \nonumber \\
&Constraints \text{  }(\ref{eq5})-(\ref{eq9}) \nonumber \hspace{127pt}\text{(I)-(V)}\\
&\alpha_{ij}-\alpha_{qj}=0, \hspace{2.55cm} j=1:v^*, \forall i,q\in \Psi \nonumber \hspace{.5cm}\text{(VI)}\\ 
&\sum_{q\in \Psi} \nolimits e_{qj}-\alpha_{ij}=0, \hspace{1.77cm} j=1:v^*,\forall i\in \Psi \nonumber \hspace{.4cm}\text{(VII)} \\ 
&\sum_{s\in \{\mathscr{S}\cup \mathscr{P}\}} \nolimits l_{si}n_{si}^j - Te_{ij} = 0,  \hspace{.6cm} i=1:v^*,\forall j\in \Psi \nonumber \hspace{.3cm}\text{(VIII)}\\
&\textit{\textbf{vars. }} \hspace{.7cm} n_{ij}^d,m_{ij}^d\geq0,0\leq \alpha_{ij}\leq \varphi_k,0\leq e_{ij}\leq 1 \nonumber
\end{align} \par
For clarification on details of the proposed models, an example for overloaded and underloaded chains is shown in Fig. \ref{fig3}. In Fig. \ref{fig3}(a), the overloaded $g_k^c$ has two VMs $(1,2)\in V$, which are running on $(\mathscr{P}_1, \mathscr{P}_2)\in \Psi$, respectively. By assuming overloaded VM 2 and setting  $v^*=4$, two extra VMs $(3,4)\in V^*$ with the same VNF type must be added  to $g_k^c$ regarding $(\mathscr{P}_2,\mathscr{P}_3, \mathscr{P}_4)\in \Psi^*$. Obtained $\alpha_{ij}$  and $e_{ij}$, VMs 3 and 4 are selected to be launched on $\mathscr{P}_3$ and $\mathscr{P}_2$, respectively. Additionally, in Fig. \ref{fig3}(b), a final decision has been made based on the calculated values of $e_{ij}$ and $\alpha_{ij}$ for the underloaded chain. In this case, the relaxed underload LP model turns off VM 3 running on $\mathscr{P}_2$.

\subsection{Introducing ADMM}

After many years without too much attention, the ADMM technique has recently witnessed a renaissance in many application areas \cite{chen2016direct}. Generally, there are two types of distributed methods: Sub-gradient and ADMM. The best known rate of convergence for sub-gradient methods is $O (\frac{1}{\sqrt{\ell}})$, while the ADMM algorithm converges at the rate $O(\frac{1}{\ell})$ for the general case, where $\ell$ is the number of iterations \cite{wei2012distributed}. Therefore, since both approaches are iteration-based and the speed of convergence in the ADMM is more than the sub-gradient method, we develop our distributed approach based on the ADMM algorithm.\par

We first introduce the background of the ADMM and then apply it to design our distributed algorithm. Consider the following convex minimization problem with a separable objective function and linear constraints:
\begin{align}
    \textit{\textbf{minimize}} \qquad &f_1(x_1)+f_2(x_2)+...+f_n(x_n), \label{eq16} \\
    \textit{\textbf{s.t.}} \qquad &A_1x_1+A_2x_2+...+A_n X_n=b, \nonumber \\
    \textit{\textbf{vars.}} \qquad &x_i \in \mathscr{X}_i, i=1,...,n, \nonumber
\end{align}
where $A_i \in \mathbb{R}^{N \times d_i}$, $b \in \mathbb{R}^{N\times1}$, $\mathscr{X}_i \subseteq \mathbb{R}^{d_i}$ is a closed convex set, and $f_i : \mathbb{R}^{d_i} \rightarrow \mathbb{R}$ is a closed convex function, $i=1,...,n$ \cite{sun2015expected}. In case of $n=2$ (i.e. two blocks of variables), the augmented Lagrangian function for model (\ref{eq16}) can be written as follows:
\begin{align}
    L_\beta(x_1,x_2,\mu)=& f_1(x_1)+f_2(x_2)+\mu^\intercal(A_1x_1+A_2x_2-b) \nonumber \\
    & +\frac{\beta}{2} \parallel A_1x_1+A_2x_2-b\parallel_2^2 \label{eq17}
\end{align}
where $\mu$ is the Lagrangian multiplier and $\beta$ is the positive penalty scalar. The two-block ADMM updates the primal variables $x_1$ and $x_2$, followed by a dual variable update $\mu$  in an iterative manner. By having $(x_1^0,x_2^0,\mu^0)$ as an initial vector, the updated variables at iteration $\ell >0$ are computed in an alternating fashion as follows:
\begin{align}
    x_1^\ell=& \argminl_{x_1} L_\beta (x_1,x_2^{\ell-1},\mu^{\ell-1}) \label{eq18}
\end{align}
\begin{align}
    x_2^\ell=&\argminl_{x_2} L_\beta (x_1^\ell,x_2,\mu^{\ell-1}) \label{eq19}
\end{align}
\begin{align}
    \mu^\ell=& \mu^{\ell-1}-\beta(A_1x_1^\ell+A_2x_2^\ell -b) \label{eq20}
\end{align}

Separable structure of convex model (\ref{eq16}) allows ADMM to decompose it over its primal variables and achieve the optimal solution in a distributed way. To formulate the ADMM into a scaled form that is more convenient to be used in designing algorithm, we define $u=(\frac{1}{\beta})\mu$ as the scaled dual variable. Thus, we have:

\begin{align}
    \mu^\intercal (A_1 & x_1+ A_2x_2-b) + \frac{\beta}{2} \parallel A_1x_1+A_2x_2-b \parallel_2^2 \label{eq21} \nonumber \\
    &= \frac{\beta}{2} \parallel A_1x_1+A_2x_2-b+u \parallel_2^2 - \frac{\beta}{2} \parallel u \parallel_2^2
\end{align}

\noindent
By disregarding independent terms of the minimization variables, the ADMM method could be expressed as: 
\begin{align}
    x_1^\ell = \argminl_{x_1} (f_1(x_1)  &+ (\frac{\beta}{2}) \parallel A_1x_1 +A_2x_2^{\ell-1} \nonumber \\
    &-b + u^{\ell-1} \parallel_2^2) \label{eq22}
\end{align}
\begin{align}
    x_2^\ell = \argminl_{x_2} (f_2(x_2) &+(\frac{\beta}{2}) \parallel A_1x_1^\ell +A_2x_2 \nonumber \\
    & -b + u^{\ell-1} \parallel_2^2 )  \label{eq23}
\end{align}
\begin{align}
    u^\ell = u^{\ell-1} + A_1 x_1^\ell + A_2 x_2^\ell - b \label{eq24}
\end{align}

The optimality and convergence of the two-block ADMM for model (\ref{eq16}) is proved in \cite{boyd2011distributed} under mild technical assumptions. 
\par On the other hand, many applications need more than two blocks of variables to be considered (i.e. $n>2$). By direct extending of ADMM to multi-block case, the convergence cannot be established well \cite{chen2016direct}. The authors in \cite{sun2015expected} proposed Randomly Permuted ADMM (RP-ADMM), as a simple modification in updating process of primal variables to make the convergence possible in a desirable duration. In fact, RP-ADMM draw a permutation $\Omega$ of $\{1,...,n\}$ uniformly at random and update the primal variables in the order of the permutation, followed by updating the dual variables in a usual way \cite{sun2015expected}. Since our proposed approach has more than two blocks of variables, we apply RP-ADMM technique to solve the problem in a distributed way.

\subsection{Applying RP-ADMM}

In this section, we develop our distributed algorithm based on the proposed model (\ref{eq14}). We firstly reformulate the proposed model (\ref{eq14}) to an applicable model for applying RP-ADMM, and then introduce our RP-ADMM-based distributed algorithm.

\par Since $\forall j \in \{ \mathscr{S} \cup \Psi \cup \Psi^* \}$, each node $i\in \{ \mathscr{S} \cup \Psi \cup \Psi^* \}$  is responsible to determine its local flow variables $n_{ij}^d$  and $m_{ij}^d$ in a distributed manner, we have to reformulate flow conservation constraints (I)-(V) in (\ref{eq14}). So, by introducing auxiliary variables $A_{ij}^d$, $B_{ij}^d$, and $C_{ij}^d$, we reform constraints (I)-(III), $j\in \{ \mathscr{S} \cup \Psi \cup \Psi^* \}, d = 1:v^*$, where $l_{ij}=1$, as follows:

\begin{align}
    &n_{ij}^d - n_{ji}^d - A_{ij}^d = 0, \hspace{.95cm} \forall i \in \mathscr{S} \label{eq25} \\
    &\sum_{j \in \{ \mathscr{S} \cup \Psi \cup \Psi^* \}} \nolimits A_{ij}^d = 0 \label{eq260} \\
    &m_{ij}^d - m_{ji}^d - B_{ij}^d  = 0, \hspace{.75cm} \forall i \in \mathscr{S} \label{eq26}\\
    &\sum_{j \in \{ \mathscr{S} \cup \Psi \cup \Psi^* \}} \nolimits B_{ij}^d = 0 \label{eq270} \\
    & m_{ij}^d - \gamma_k n_{ji}^d - C_{ij}^d  =0 , \hspace{.52cm} \forall i \in \mathscr{P} \label{eq27} \\
    &\sum_{j \in \{ \mathscr{S} \cup \Psi \cup \Psi^* \}} \nolimits C_{ij}^d = 0 \label{eq300}
\end{align}

Without loss of generality, assume that ingress and egress points of $g_k^c$ have one PM indexed by 0 and 1, respectively. To reform constraints (IV) and (V) in (\ref{eq14}), we similarly define two more auxiliary variables $D_d$ and $E_d$, $(d=1:v^*)$:
\begin{align}
    &\sum_{j \in \{\mathscr{S} \cup \mathscr{P}_0\}} \nolimits l_{0j} n_{0j}^d - D_d = 0 \label{eq28} \\
    &\sum_{d=1:v^*} \nolimits D_d - T = 0 \label{eq280} \\
    &\sum_{j \in \{\mathscr{S} \cup \mathscr{P}_1\}} \nolimits l_{j1} m_{j1}^d - E_d = 0\label{eq29}\\
    &\sum_{d=1:v^*} \nolimits E_d - T\gamma_k = 0 \label{eq340}
\end{align}
\noindent
Now, the RP-ADMM can be easily applied on the following model:
\begin{align}
    \textit{\textbf{minimize}} \qquad &Eq. (\ref{eq12}) \label{eq350} \\
    \textit{\textbf{s.t.}} \qquad &Eq.(\ref{eq25}) \text{ - } Eq.(\ref{eq340}), \nonumber \\
    & \textit{constraints (VI) - (IX) in model (\ref{eq14})} \nonumber \\
    \textit{\textbf{vars.}} \qquad &n_{ij}^d,m_{ij}^d, A_{ij}^d,B_{ij}^d,C_{ij}^d,D_d,E_d\geq0, \nonumber \\
    & 0\leq \alpha_{ij}\leq \varphi_k,0\leq e_{ij}\leq 1 \nonumber
\end{align}

The augmented Lagrangian function for the model (\ref{eq350}) can be presented as follows:
\begin{align}
    &L_\beta (n,m,A,B,C,D,E,\alpha,e; \label{eqlag} \\
    &\delta,\bar{\delta}, \eta, \bar{\eta}, \zeta , \bar{\zeta}, \lambda_0,\bar{\lambda}_0, \lambda_1, \bar{\lambda}_1, \mu ,\xi , \sigma , \rho) = Eq. (\ref{eq12}) \nonumber \\
    & + \delta^\intercal Eq.(\ref{eq25}) + \frac{\beta}{2} \parallel Eq. (\ref{eq25}) \parallel_2^2 
    + \bar{\delta}^\intercal Eq.(\ref{eq260}) + \frac{\beta}{2} \parallel Eq. (\ref{eq260}) \parallel_2^2 \nonumber \\
    & + \eta^\intercal Eq.(\ref{eq26}) + \frac{\beta}{2} \parallel Eq. (\ref{eq26}) \parallel_2^2  + \bar{\eta}^\intercal Eq.(\ref{eq270}) + \frac{\beta}{2} \parallel Eq. (\ref{eq270}) \parallel_2^2 \nonumber \\
    & + \zeta^\intercal Eq.(\ref{eq27}) + \frac{\beta}{2} \parallel Eq. (\ref{eq27}) \parallel_2^2  + \bar{\zeta}^\intercal Eq.(\ref{eq300}) + \frac{\beta}{2} \parallel Eq. (\ref{eq300}) \parallel_2^2 \nonumber \\
    & + \lambda_0^\intercal Eq.(\ref{eq28}) + \frac{\beta}{2} \parallel Eq. (\ref{eq28}) \parallel_2^2  + \bar{\lambda}_0^\intercal Eq.(\ref{eq280}) + \frac{\beta}{2} \parallel Eq. (\ref{eq280}) \parallel_2^2 \nonumber \\
    & + \lambda_1^\intercal Eq.(\ref{eq29}) + \frac{\beta}{2} \parallel Eq. (\ref{eq29}) \parallel_2^2  + \bar{\lambda}_1^\intercal Eq.(\ref{eq340}) + \frac{\beta}{2} \parallel Eq. (\ref{eq340}) \parallel_2^2 \nonumber \\
    & + \mu^\intercal \text{Cons. (VI, model (\ref{eq14}))} + \frac{\beta}{2} \parallel \text{Cons. (VI model (\ref{eq14}))} \parallel_2^2 \nonumber \\
    & + \xi^\intercal \text{Cons. (VII, model (\ref{eq14}))} + \frac{\beta}{2} \parallel \text{Cons. (VII model (\ref{eq14}))} \parallel_2^2 \nonumber \\
    & + \sigma^\intercal \text{Cons. (VIII, model (\ref{eq14}))} + \frac{\beta}{2} \parallel \text{Cons. (VIII model (\ref{eq14}))} \parallel_2^2 \nonumber \\
    & + \rho^\intercal \text{Cons. (IX, model (\ref{eq14}))} + \frac{\beta}{2} \parallel \text{Cons. (IX model (\ref{eq14}))} \parallel_2^2 \nonumber
\end{align}
\noindent where $\delta,\bar{\delta}, \eta, \bar{\eta}, \zeta , \bar{\zeta}, \lambda_0,\bar{\lambda}_0, \lambda_1, \bar{\lambda}_1, \mu ,\xi , \sigma$, and $\rho$ are the Lagrangian multiplier and $\beta$ is the positivie penalty scalar.
\par Considering the above explanations, we are able to utilize the \textit{n}-block RP-ADMM algorithm to distributedly solve the model (\ref{eq350}) (see Algorithm \ref{alg1}). As illustrated in Algorithm \ref{alg1}, it starts by initializing primal and dual variables in the first line. In line 2, we defined an array to store the last values of the variables (primal and dual). The iterative process begins in line 3 where $\ell$ indicates the round number. In each round, RP-ADMM algorithm picks a uniformly random permutation $\Omega$ to specify the order of updating primal variables. In fact, the inner $for$ loop (lines 6 to 10), selects the $\Omega(i)^{th}$ variable to update its value in $\ell^{th}$ round by considering the last value of variables stored in the \textit{LastValueArray}. Then, in line 9 the value of $\Omega(i)^{th}$ primal variable is overwrited. For illustration, suppose that $\Omega = \{3,4,2,1\}$ ($n=4$) and $\ell = 12$. Therefore, according to the order of $\Omega$, the third primal variable (i.e. $\Omega(1)^{th}$ variable) must be updated first, followed by fourth, second, and first primal variable. Also, in this round, for example, the second variable updating process uses the current values (computed in 12$^{th}$ round) of third and forth variables, and also the obtained value of the first variable in previous round (i.e. 11$^{th}$ round). Using the scaled form of ADMM algorithm, the primal variables are updated as follows:
\par \textit{\textbf{Updating Switches Primal Variables:}} If the selected variable belongs to the set of network switches, each switch $i$, updates its traffic flow rate variable to its neighbor $j\in \{\mathscr{S}\cup \Psi \cup \Psi^*\}$ using the following equations:
\begin{align}
    &n_{ij}^{d(\ell)} = \argminl_{n_{ij}^d} (Eq. (\ref{eqlag})) \nonumber \\
    & \hspace{.7cm}= \argminl_{n_{ij}^d}  (Eq.(\ref{eq12})+\frac{\beta}{2} \parallel n_{ij}^d - n_{ji}^{d(\ell^*)} - A_{ij}^{d(\ell^*)} \nonumber \\
    & + \frac{1}{\beta} \delta_{ij}^{d(\ell^*)} \parallel_2^2  + \frac{\beta}{2}  \parallel n_{ji}^{d(\ell^*)}  -  n_{ij}^d  - A_{ji}^{d(\ell^*)}   +  \frac{1}{\beta} \delta_{ji}^{d(\ell^*)} \parallel_2^2 \nonumber \\
    &+ \frac{\beta}{2} \parallel m_{ji}^{d(\ell^*)}  - \gamma_k n_{ij}^d - C_{ji}^{d(\ell^*)} + \frac{1}{\beta} \zeta_{ji}^{d(\ell^*)} \parallel_2^2 + \frac{\beta}{2} \parallel  n_{ij}^d  \nonumber \\
    & +  \sum_{s\in \{\mathscr{S}\cup \Psi \cup \Psi^*\}, s \neq i} \nolimits l_{sj} n_{sj}^{d(\ell^*)}  -  T e_{jd}^{\ell^*} + \frac{1}{\beta} \sigma_j^{d(\ell^*)} \parallel_2^2     \nonumber \\
    &+   \argminr_{j\in \Psi}  \frac{\beta}{2} \parallel n_{ij}^d + \sum_{s\in \{\mathscr{S} \cup \Psi \cup \Psi^*\}, s \neq i} \nolimits l_{sj} n_{sj}^{d(\ell^*)}   -   T \alpha_{jd}^{(\ell^*)} \nonumber \\
    &+ \frac{1}{\beta} \rho_j^{d(\ell^*)}  \parallel_2^2    ) \label{eq30} 
\end{align}
where $\argminr_{X}$ plays as an if statement; i.e. it returns 1 if condition \textit{"X"} is true. Also, $a^{d(\ell^*)}$ or $a^{(\ell^*)}$ shows the most recent value of the primal variable $a$, stored in \textit{LastValueArray}. Similarly, $m_{ij}^{d(\ell)}$, $A_{ij}^{d(\ell)}$, and $B_{ij}^{d(\ell)}$ are obtained from the following equations: 
\begin{align}
    &m_{ij}^{d(\ell)} = \argminl_{m_{ij}^d}  ( Eq. (\ref{eq12}) + \frac{\beta}{2} \parallel m_{ij}^d - m_{ji}^{d(\ell^*)} - B_{ij}^{d(\ell^*)} \nonumber \\ 
    &+ \frac{1}{\beta} \eta_{ij}^{d(\ell^*)} \parallel_2^2  + \frac{\beta}{2} \parallel m_{ji}^{d(\ell^*)}  - m_{ij}^d - B_{ij}^{d(\ell^*)} + \frac{1}{\beta} \eta_{ji}^{d(\ell^*)} \parallel_2^2 \nonumber \\
    &+ \argminr_{j=1}  \frac{\beta}{2} \parallel m_{ij}^d + \sum_{s\in \{\mathscr{S} \cup \mathscr{P}_1 \}, s \neq i } \nolimits l_{sj} m_{sj}^{d(\ell^*)}  -  E_d^{(\ell^*)} \nonumber \\
    & + \frac{1}{\beta} \lambda_j^{d(\ell^*)} \parallel_2^2  ) \label{eq31}
\end{align}
\begin{align}
    & A_{ij}^{d(\ell)}  =  \argminl_{A_{ij}^d}   ( \frac{\beta}{2} \parallel n_{ij}^{d(\ell^*)} - n_{ji}^{d(\ell^*)} - A_{ij}^d + \frac{1}{\beta} \delta_{ij}^{d(\ell^*)} \parallel_2^2 \nonumber \\
    & +\frac{\beta}{2} \parallel A_{ij}^d + \sum_{s\in \{\mathscr{S} \cup \Psi \cup \Psi^*\},s \neq j} \nolimits A_{is}^{d(\ell^*)}  + \frac{1}{\beta} \bar{\delta}_i^{d(\ell^*)} \parallel_2^2  ) \label{eq37}
\end{align}
\begin{align}
    & B_{ij}^{d(\ell)}  =  \argminl_{B_{ij}^d}   ( \frac{\beta}{2} \parallel m_{ij}^{d(\ell^*)} - m_{ji}^{d(\ell^*)} - B_{ij}^d + \frac{1}{\beta} \eta_{ij}^{d(\ell^*)} \parallel_2^2 \nonumber \\
    & +\frac{\beta}{2} \parallel B_{ij}^d + \sum_{s\in \{\mathscr{S} \cup \Psi \cup \Psi^*\},s \neq j} \nolimits B_{is}^{d(\ell^*)}  + \frac{1}{\beta} \bar{\eta}_i^{d(\ell^*)} \parallel_2^2  ) \label{eq38}
\end{align}

\begin{algorithm}[t]
\caption{\textit{n}-block RP-ADMM}
\label{alg1}
\SetAlgoLined
Initialize: {$n_{ij}^{d(0)}$, $m_{ij}^{d(0)}$, $\alpha_{ij}^{d(0)}$, $e_{ij}^{d(0)}$, $A_{ij}^{d(0)}$, $B_{ij}^{d(0)}$, $C_{ij}^{d(0)}$, $D_d^{(0)}$, $E_d^{(0)}$ and dual variables}\\
 LastValueArray[] = \textit{initialized variables}\;
 \For {$\ell =1,2,...$} {
    $n$=length(primal variables)\;
    Pick a permutation $\Omega$ of $\{1,...,n\}$ uniformly at random\;
    \For {$i =1 :n$} { 
        //Updating primal variables:\\
        Update the $\Omega(i)^{th}$ primal variable\;
        Update LastValueArray[$\Omega(i)$]\;   
    }
    Update dual variables\;
 }
\end{algorithm}

\par \textit{\textbf{Updating PMs Primal Variables:}} If the selected variable belongs to the set of PMs, each PM $i$ updates the primal dual variables $n_{ij}^d$, $m_{ij}^d$, $\alpha_{id}$, $e_{id}$, $C_{ij}^{d(\ell)}$, $D_{d}^{(\ell)}$, and $E_{d}^{\ell}$  using the following equations:
\begin{align}
    & n_{ij}^{d(\ell)} = \argminl_{n_{ij}^d}  ( Eq.(\ref{eq12}) + \frac{\beta}{2} \parallel n_{ji}^{d(\ell^*)} - n_{ij}^d - A_{ji}^{d(\ell^*)}  \nonumber \\
    & +  \frac{1}{\beta} \delta_{ji}^{d(\ell^*)} \parallel_2^2 + \argminr_{i=0} \frac{\beta}{2} \parallel n_{ij}^d + \sum_{s\in \{\mathscr{S} \cup \mathscr{P}_0\}, s \neq j} \nolimits l_{is} n_{is}^{d(\ell^*)} \nonumber \\ 
    & - D_d^{(\ell^*)} + \frac{1}{\beta} \lambda_0^{d(\ell^*)} \parallel_2^2 + \frac{\beta}{2} \parallel n_{ij}^d  \nonumber \\ 
    & + \sum_{s\in \{\mathscr{S} \cup \Psi \cup \Psi^*\},s \neq i} \nolimits l_{sj} n_{sj}^{d(\ell^*)}- T e_{jd}^{d(\ell^*)} + \frac{1}{\beta} \sigma_j^{d(\ell^*)} \parallel_2^2  \nonumber \\
    & + \frac{\beta}{2} \parallel n_{ij}^d + \sum_{s\in \{\mathscr{S} \cup \Psi \cup \Psi^*\},s \neq i} \nolimits l_{sj} n_{sj}^{d(\ell^*)}- T \alpha_{jd}^{d(\ell^*)} 
    \nonumber \\
    & + \frac{1}{\beta} \rho_j^{d(\ell^*)}  \parallel_2^2  ) \label{eq32}
\end{align}
\begin{align}
    & m_{ij}^{d(\ell)} = \argminl_{m_{ij}^d}  ( Eq.(\ref{eq12}) + \frac{\beta}{2} \parallel m_{ji}^{d(\ell^*)} - m_{ij}^d - B_{ji}^{d(\ell^*)}  \nonumber \\
    & +  \frac{1}{\beta} \eta_{ji}^{d(\ell^*)} \parallel_2^2 + \frac{\beta}{2} \parallel m_{ij}^d - \gamma_k n_{ij}^{d(\ell^*)} - C_{ij}^{d(\ell^*)} + \frac{1}{\beta} \zeta_{ij}^{d(\ell^*)} \parallel_2^2 \nonumber \\
    & + \argminr_{i=1} \frac{\beta}{2} \parallel m_{ij}^d + \sum_{s\in \{ \mathscr{S} \cup \mathscr{P}_1 \}, s \neq i} \nolimits l_{sj} m_{sj}^{d(\ell^*)} - E_d^{(\ell^*)} \nonumber \\
    & +\frac{1}{\beta} \lambda_i^{d(\ell^*)} \parallel_2^2  ) \label{eq33} 
\end{align}
\begin{align}
    & \alpha_{id}^{(\ell )} = \argminl_{\alpha_{id}}  ( \argminr_{d>|V|} \frac{\beta}{2} \sum_{q\in \{\Psi \cup \Psi^*\} , q \neq i} \nolimits \parallel \alpha_{id} - \alpha_{qd}^{(\ell^*)} \nonumber \\
    & + \frac{1}{\beta} \mu_{iq}^{d(\ell^*)} \parallel_2^2 + \frac{\beta}{2} \parallel \sum_{q\in \{\Psi \cup \Psi^* \}} \nolimits e_{qd}^{(\ell^*)} - \alpha_{id} + \frac{1}{\beta} \xi_i^{d(\ell^*)} \parallel_2^2 \nonumber \\
    & + \argminr_{i \in \Psi} \frac{\beta}{2} \parallel \sum_{s\in \{\mathscr{S} \cup \Psi \cup \Psi^*\}} \nolimits l_{si} n_{si}^{d(\ell^*)} - T \alpha_{id} + \frac{1}{\beta} \rho_i^{d(\ell^*)} \parallel_2^2  ) \label{eq34}
\end{align}
\begin{align}
    & e_{id}^{(\ell)} = \argminl_{e_{id}}  ( \frac{\beta}{2} \parallel e_{id} + \sum_{q\in \{\Psi \cup \Psi^*\} , q \neq i} \nolimits e_{id}^{(\ell^*)} - \alpha_{qd}^{(\ell^*)} \nonumber \\ 
    & + \frac{1}{\beta} \xi_i^{d(\ell^*)} \parallel_2^2 + \argminr_{d>|V} \frac{\beta}{2} \parallel \sum_{s\in \{\mathscr{S} \cup \Psi \cup \Psi^*\}} \nolimits l_{si} n_{si}^{d(\ell^*)} - T e_{id} \nonumber \\
    & + \frac{1}{\beta} \sigma_i^{d(\ell^*)} \parallel_2^2  ) \label{eq35}
\end{align}
\begin{align}
    & C_{ij}^{d(\ell)}  =  \argminl_{C_{ij}^d}   ( \frac{\beta}{2} \parallel m_{ij}^{d(\ell^*)} - \gamma_k n_{ji}^{d(\ell^*)} - C_{ij}^d + \frac{1}{\beta} \zeta_{ij}^{d(\ell^*)} \parallel_2^2 \nonumber \\
    & +\frac{\beta}{2} \parallel C_{ij}^d + \sum_{s\in \{\mathscr{S} \cup \Psi \cup \Psi^*\},s \neq j} \nolimits C_{is}^{d(\ell^*)}  +\frac{1}{\beta} \bar{\zeta}_i^{d(\ell^*)} \parallel_2^2  ) \label{eq39}
\end{align} 

In case of  $i \in ingress$, we should update $D_d^{(\ell)}$ as below:

\begin{align}
    &D_d^{(\ell)} = \argminl_{D_d}  ( \frac{\beta}{2}  ( \parallel  \sum_{s \in \{\mathscr{S} \cup \mathscr{P}_0\}} \nolimits l_{is} n_{is}^{d(\ell^*)} - D_d \nonumber \\
    &+ \frac{1}{\beta} \lambda_i^{d(\ell^*)} \parallel_2^2 + \parallel D_d + \sum_{s=1:v^* , s \neq d} \nolimits D_s^{(\ell^*)} - T \nonumber \\
    &+ \frac{1}{\beta} \bar{\lambda}_i^{(\ell^*)} \parallel_2^2  )  ) \label{eq40}
\end{align}
However, if $i \in egress$, $E_d^{(\ell)}$ should be updated as follows:
\begin{align}
    & E_d^{(\ell)} = \argminl_{E_d}  ( \frac{\beta}{2}  ( \parallel  \sum_{s \in \{\mathscr{S} \cup \mathscr{P}_1\}} \nolimits l_{si} m_{si}^{d(\ell^*)} - E_d \nonumber \\
    &+ \frac{1}{\beta} \lambda_i^{d(\ell^*)} \parallel_2^2 + \parallel E_d + \sum_{s=1:v^* , s \neq d} \nolimits E_s^{(\ell^*)} - T \nonumber \\
    &+ \frac{1}{\beta} \bar{\lambda}_i^{(\ell^*)} \parallel_2^2  )  ) \label{eq41}
\end{align}
Finally, in line 11 of Algorithm \ref{alg1}, the dual variables can be updated as follows:
\begin{itemize}
  \item $\forall i \in \mathscr{S}$ and $  \forall j \in \{\mathscr{S} \cup \Psi \cup \Psi^*\}:$
\end{itemize}
\begin{align}
    & \delta_{ij}^{d(\ell )} = \delta_{ij}^{d(\ell-1)} + \beta( n_{ij}^{d(\ell )} - n_{ji}^{d(\ell)} - A_{ij}^{d(\ell)} ) \\
    &\bar{\delta_i}^{d(\ell )} = \bar{\delta_i}^{d(\ell-1)} + \beta( \sum_{s \in \{\mathscr{S} \cup \mathscr{P}\}} \nolimits A_{is}^{d(\ell)} ) \\
    &\eta_{ij}^{d(\ell) } = \eta_{ij}^{d(\ell-1)} + \beta( m_{ij}^{d(\ell )} - m_{ji}^{d(\ell )} - B_{ij}^{d(\ell )} ) \\
    &\bar{\eta_i}^{d(\ell)} = \bar{\eta_i}^{d(\ell-1)} + \beta( \sum_{s \in \{\mathscr{S} \cup \mathscr{P}\}} \nolimits B_{is}^{d(\ell)} ) 
\end{align}
\begin{itemize}
  \item $\forall i \in \{\Psi \cup \Psi^*\}\text{ and }\forall j \in \{\mathscr{S} \cup \Psi \cup \Psi^*\}:$
\end{itemize}

\begin{align}
    &\zeta_{ij}^{d(\ell)} = \zeta_{ij}^{d(\ell-1)} + \beta( m_{ij}^{d(\ell)} - \gamma_k n_{ji}^{d(\ell )} - C_{ij}^{d(\ell)} )
\end{align}
\begin{align}
    &\bar{\zeta_i}^{d(\ell )} = \bar{\zeta_i}^{d(\ell-1)} + \beta( \sum_{s \in \{\mathscr{S} \cup \mathscr{P}\}} \nolimits C_{il}^{d(\ell)} ) \\
    &\sigma_i^{d(\ell)} = \sigma_i^{d(\ell-1)} + \beta (\sum_{s\in \{\mathscr{S} \cup \mathscr{P}\}} \nolimits l_{si} n_{si}^{d(\ell)} -  T e_{id}^{(\ell)} )\\
    &\mu_{ij}^{d(\ell)} = \mu_{ij}^{d(\ell-1)} + \beta (\alpha_{id}^{(\ell)} - \alpha_{jd}^{(\ell)}  ) \\
    &\xi_i^{d(\ell)} = \xi_i^{d(\ell-1)} + \beta (\sum_{l\in \{\Psi \cup \Psi^*\}} \nolimits  e_{ld}^{d(\ell)} - \alpha_{id}^{(\ell)}    )
\end{align}
\begin{itemize}
  \item If $(i \in ingress):$
\end{itemize}
\begin{align}
    &\lambda_0^{d(\ell)} = \lambda_0^{d(\ell-1)} + \beta (\sum_{s\in \{\mathscr{S} \cup \mathscr{P}_0\}} \nolimits l_{0s}n_{0s}^{d(\ell)} - D_d^{(\ell)}) \\
    &\bar{\lambda}_0^{(\ell)} = \bar{\lambda}_0^{(\ell-1)} + \beta (\sum_{d=1:v^*} \nolimits D_d^{(\ell)})
\end{align}
\begin{itemize}
  \item If $(i \in egress):$
\end{itemize}
\begin{align}
    &\lambda_1^{d(\ell)} = \lambda_1^{d(\ell-1)} + \beta (\sum_{s\in \{\mathscr{S} \cup \mathscr{P}_1\}} \nolimits l_{s1} m_{s1}^{d(\ell)} - E_d^{(\ell)}) \\
    &\bar{\lambda}_1^{(\ell)} = \bar{\lambda_1}^{(\ell-1)} + \beta (\sum_{d=1:v^*} \nolimits E_d^{(\ell)})
\end{align}
\begin{itemize}
  \item If $(i \in \Psi):$
\end{itemize}
\begin{align}
    \rho_i^{d(\ell)} = \rho_i^{d(\ell-1)} + \beta (\sum_{s\in \{\mathscr{S} \cup \mathscr{P}\}} \nolimits l_{si} n_{si}^{d(\ell)} -  T \alpha_{id}^{(\ell)} )
\end{align}

\begin{table*}[t]
\centering
\caption{Comparing time complexities of proposed models in different topologies}
\label{table2}
\begin{tabular}{cccccccccc}
\toprule
& & & & \multicolumn{2}{c}{MILP (\ref{eq13})} & \multicolumn{2}{c}{LP (Overload) (\ref{eq14})} & \multicolumn{2}{c}{LP (Underload) (\ref{eq15})} \\ 
\cmidrule(r){5-6}
\cmidrule(r){7-8}
\cmidrule(r){9-10}
$k$-fat-tree & No. of Switches & No. of PMs & No. of Variables & No. of Constraints & Run time (s) & No. of Constraints  & Run time (s) & No. of Constraints & Run Time (s) \\ \hline 
\textbf{2} & 5 & 4 & 120 & 25 & 0.555 & 68 & 0.075 & 62 & 0.055 \\ \hline
\textbf{4} & 20 & 16 & 672 & 315 & 5.064 & 254 & 0.614 & 188 & 0.318 \\ \hline
\textbf{8} & 80 & 64 & 4224 & 1251 & 323.523 & 998 & 14.977 & 692 & 6.635 \\ \hline
\textbf{16} & 320 & 256 & 29184 & 4995 & $N/A$ & 3974 & 974.235 & 2708 & 456.674 \\ \hline
\textbf{32} & 1280 & 1024 & 129024 & 19971 & $N/A$ & 15878 & $N/A$ & 10772 & $N/A$ \\ \hline
\textbf{64} & 5120 & 4096 & 466944 & 79875 & $N/A$ & 63494 & $N/A$ & 43028 & $N/A$ \\
\bottomrule
\end{tabular}
\end{table*}

\par Clearly, the primal values of the RP-ADMM model can be updated according to the selected permutation. In case of overload, when  $v_k^{i,c}$ violates the defined threshold, it determines $v^*$ and then triggers PMs $\in \{\Psi \cup \Psi^*\}$ and some adjacent switches via sending a simple packet. Then, the RP-ADMM update process can be launched considering the selected PMs and switches.

\section{Performance Evaluation}

\begin{figure}[t]
\centering
\includegraphics[scale=1]{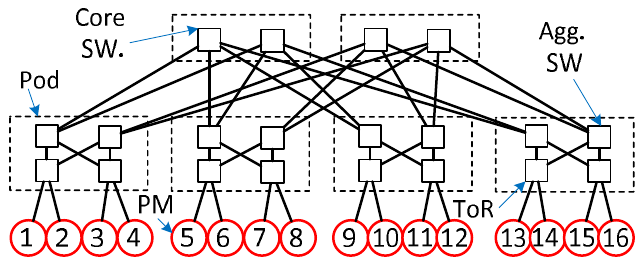}
\caption{4-fat-tree topology}
\label{fig5}
\end{figure}

To assess the performance of the proposed approach, we simulate the proposed MILP, LP, and ADMM models utilizing MATLAB software. We conducted our simulation on a hardware composing of an INTEL Core i7 1.73GHz CPU, 8GB of RAM, and Windows 10 x64 OS. We considered $k$-fat-tree topologies as the datacenter network structure. In general, a $k-fat-tree$ has $k/2$ ToR and aggregate switches in each pod, and $k^2/4$ core switches \cite{tammana2015cherrypick}. We mainly use 4-ary fat-tree topology for performance evaluation (see Fig. \ref{fig5}). However, we run some of our experiments based on different topology sizes (i.e. 8, 16, 32, and 64-ary fat-trees). Also $\tau$ is considered as 5 seconds for all runs. \par

In our first experiment, we investigate the performances of MILP (\ref{eq13}), LP (\ref{eq14}), and (\ref{eq15}) models in terms of the number of variables, constraints, and execution time on various k-fat-trees. We consider $g_c^k$ with two VMs (i.e. $|V|=2$), $|\Psi|=2$, $|V^*|=1$, and each rack equipped with two PMs. As it can be seen in Table \ref{table2}, number of switches, PMs, variables, and constraints are proportional to the size of datacenter (i.e. $k$-fat-tree). Obviously, for larger datacenters, the execution time of the proposed models also increased specifically for MILP model. Table \ref{table2} clearly implies that in LDCs, we need to apply an efficient distributed algorithm to achieve a solution in a reasonable time. 

\par As another experiment, we consider a 4-fat-tree topology to evaluate the performance of the proposed LP models for over and underloaded chains in different scenarios. Each of these scenarios is designed based on the number of PMs in $g_k^c$, ingress and egress points, $\Psi^*$ (i.e. the candidate PMs that are able to launch new VMs), normalized ($0 \sim 1$) forwarding costs (N. FwdCost), and $v^*$ (the number of required VNFs to process the received traffic by $g_k^c$) for overload and underload cases (see column 1-4 in Table \ref{table3}). We also considered the forwarding costs between different layers of datacenter as: PM-ToR=10, ToR-Aggregation=20, and Aggregation-Core=40. Further, to simulate overload and underload cases, we increased the traffic rate by $50\%$ (for each extra VM) and decreased it by $50\%$, respectively. 

\par The normalized forwarding cost and its increase/decrease rates are shown in Table \ref{table3}. We also configured the initial parameters in such a way that LP is forced to turn on/off VMs to mitigate the over or underload $g_k^c$. Overall, Table \ref{table3} shows how LP models react to over/underloaded chains by turning on/off offline/online VMs. In case of overload, the proposed LP model (\ref{eq14}) prefers to launch a VM on ingress and egress PMs in order to minimize the traffic forwarding cost. In fact, the traffic between two different VNFs on an identical PM (i.e. ingress or egress) is negligible due to an internal traffic handling. For instance, in the first scenario in Table \ref{table3} ($\mathscr{P}_1 \rightarrow (\mathscr{P}_2 , \mathscr{P}_5 ) \rightarrow \mathscr{P}_4$), when we need to turn on two more VMs for $g_k^c$ (i.e. $v^* = 4$), the LP model (\ref{eq14}) selects $\mathscr{P}_1$ and $\mathscr{P}_4$ (i.e. ingress and egress points, respectively) to launch the new VMs ($\mathscr{P}_1 \rightarrow (\pmb{\mathscr{P}_1}, \mathscr{P}_2 , \mathscr{P}_5, \pmb{\mathscr{P}_4} ) \rightarrow \mathscr{P}_4$). Also, it can be seen that by using this new configuration, the forwarding cost increases by $20\%$, which is caused by the increased amount of network traffic. In addition, it can be seen that in the third scenario ($\mathscr{P}_1 \rightarrow (\mathscr{P}_1,\mathscr{P}_2) \rightarrow \mathscr{P}_2$), the maximum increase in forwarding cost is resulted among all scenarios ($116\%$). This is because, the utilized PMs before chain overload (i.e. $\mathscr{P}_1$ and $\mathscr{P}_2$) were on a same rack, while after running LP model (\ref{eq14}), two more PMs (i.e. $\mathscr{P}_3$ and $\mathscr{P}_4$) are use to handle the traffic, which are stored on another rack. It is noticeable that we assumed that only $\mathscr{P}_3 - \mathscr{P}_{16}$ PMs can afford enough resources to launch new VMs (i.e. $\Psi^* = \{\mathscr{P}_3 - \mathscr{P}_{16}\}$). On the other hand, when an underload occurs, the LP model (\ref{eq15}) prefers to turn off a VM that has the most forwarding cost. For example, in the first scenario, $\mathscr{P}_5$ is turned off. Because $\mathscr{P}_5$ is in pod 2, so the traffic should pass the higher layers of network (i.e. aggregate and core layers) to reach the next hop in the chain (i.e. $\mathscr{P}_4$). Also, as it can be seen in the last column of Table \ref{table3}, the forwarding cost decreases by $14\%$. Additionally, it can be seen that maximum forwarding cost reduction took place in the last scenario. The reason is that the LP model (\ref{eq15}) turns off the VMs on $\mathscr{P}_3$ and $\mathscr{P}_4$, which are stored in another rack. Therefore, the traffic is not passing the higher network layers and as a result, the forwarding cost in the chain reduces significantly.

\begin{figure}[t]
\centering
\includegraphics[width=0.52\linewidth]{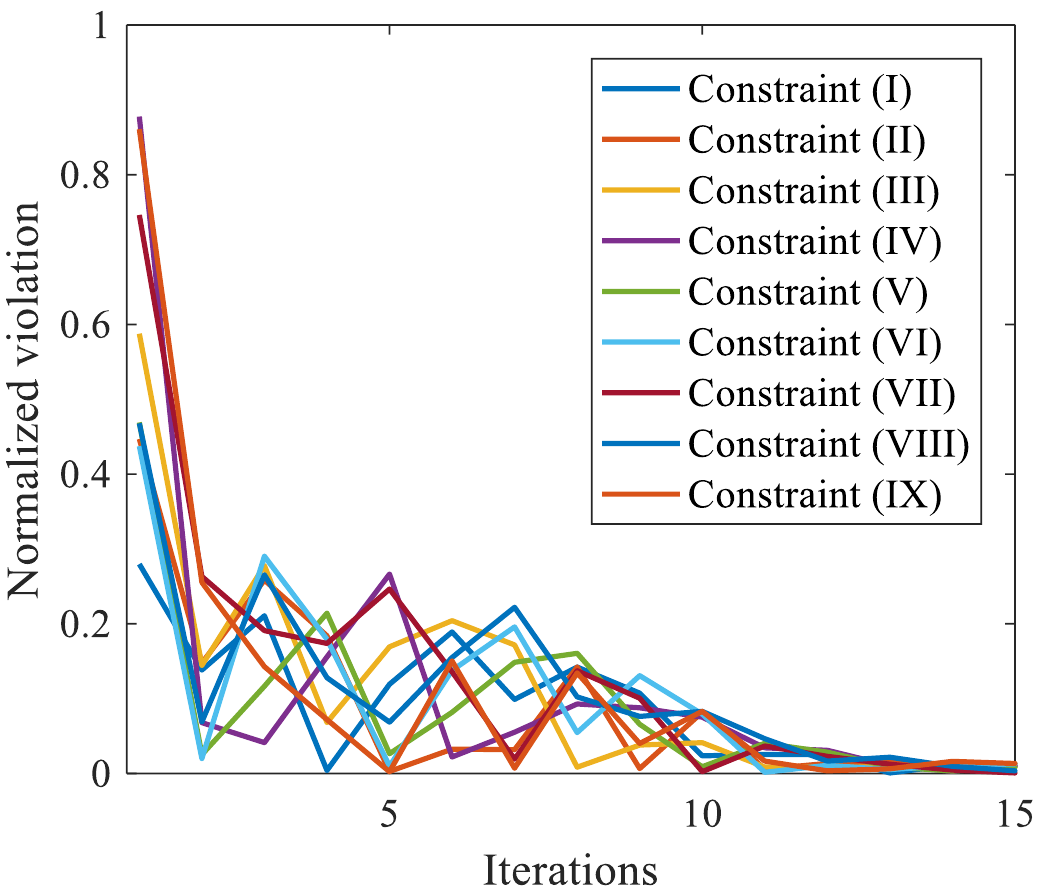}
\caption{The normalized violation of the constraints in model (\ref{eq14})}
\label{fig7}
\end{figure}

\begin{figure}[t]
\centering
\includegraphics[width=0.52\linewidth]{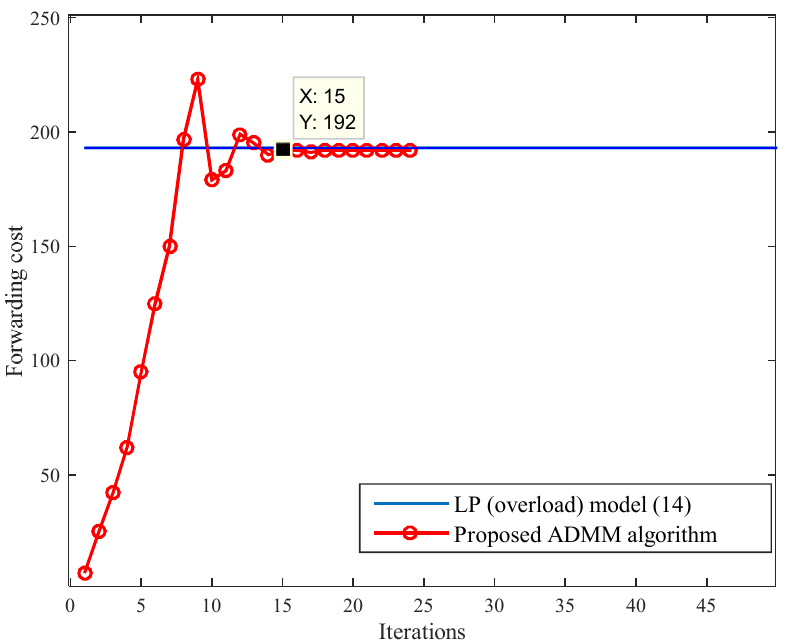}
\caption{The convergence rate of the proposed distributed RP-ADMM algorithm}
\label{fig6}
\end{figure}

One of the well-known features of ADMM is its fast convergence which is clearly illustrated in Fig. \ref{fig6}. For declaration, we focused on the last scenario in Table (\ref{table3}) and compared the optimal cost value obtained by LP model (\ref{eq14}) with the solution of distributed ADMM algorithm (we set $\beta =5$). According to Fig. \ref{fig6}, the distributed ADMM converges to the optimal value in 15 iterations. This low number of iterations indicates that the proposed distributed ADMM technique can perform well with low overhead (e.g. message passing) in a LDC. Notably, we set the stop criterion to 25 iterations. In addition, in Fig. \ref{fig7} we measure the normalized violation of each constraint in model (\ref{eq14}) to show how fast violations decrease.

\begin{table*}[t!]
\centering
\caption{Comparing the results of LP models in different scenarios}
\label{table3} 
\begin{tabular}{|c|c|c|c|c|c|c|c|c|}

\hline

\multirow{2}{*}{$g_{k-1}^c \rightarrow (g_k^c) \rightarrow g_{k+1}^c$} & \multirow{2}{*}{N. FwdCost} & \multicolumn{4}{c|}{LP overload model (\ref{eq14})} & \multicolumn{3}{c|}{LP underload model (\ref{eq15})}

\\ \cline{3-9}

& & $\Psi^*$ & $v^*$ & New Configuration &  FwdCost & $v^*$ & New Configuration &  FwdCost \\ \hline

\multirow{2}{*}{$\mathscr{P}_1 \rightarrow (\mathscr{P}_2 , \mathscr{P}_5) \rightarrow \mathscr{P}_4 $} & \multirow{2}{*}{0.68} & \multirow{2}{*}{ALL PMs} & 3 & $\mathscr{P}_1 \rightarrow (\mathscr{P}_2 , \mathscr{P}_5 , \pmb{\mathscr{P}_4}) \rightarrow \mathscr{P}_4$ & +14\% & \multirow{2}{*}{1} & \multirow{2}{*}{$\mathscr{P}_1 \rightarrow (\mathscr{P}_2) \rightarrow \mathscr{P}_4$} & \multirow{2}{*}{-14\%} \\ \cline{4-6}
& & & 4 & $\mathscr{P}_1 \rightarrow (\pmb{\mathscr{P}_1}, \mathscr{P}_2,\mathscr{P}_5, \pmb{\mathscr{P}_4} ) \rightarrow \mathscr{P}_4$ & +20\% & & & \\ \hline

\multirow{2}{*}{$\mathscr{P}_1 \rightarrow ( \mathscr{P}_1,\mathscr{P}_2,\mathscr{P}_3) \rightarrow \mathscr{P}_{16}$} & \multirow{2}{*}{1} & \multirow{2}{*}{$\mathscr{P}_3 - \mathscr{P}_{16}$} & 4 & $\mathscr{P}_1 \rightarrow (\mathscr{P}_1, \mathscr{P}_2,\mathscr{P}_3,\pmb{\mathscr{P}_{16}}) \rightarrow \mathscr{P}_{16}$ & +11\% & 2 & $\mathscr{P}_1 \rightarrow (\mathscr{P}_1,\mathscr{P}_2) \rightarrow \mathscr{P}_4$ & -6\% \\ \cline{4-9}
& & & 5 & $\mathscr{P}_1 \rightarrow (\mathscr{P}_1,\mathscr{P}_2,\mathscr{P}_3,\pmb{\mathscr{P}_3},\pmb{\mathscr{P}_{16}}) \rightarrow \mathscr{P}_{16}$ & +17\% & 1 & $\mathscr{P}_1 \rightarrow (\mathscr{P}_1) \rightarrow \mathscr{P}_4 $ & -16\% \\ \hline

\multirow{2}{*}{$\mathscr{P}_1 \rightarrow (\mathscr{P}_1,\mathscr{P}_2) \rightarrow \mathscr{P}_2$} & \multirow{2}{*}{0.23} & ALL PMs & 3 & $\mathscr{P}_1 \rightarrow (\mathscr{P}_1,\pmb{\mathscr{P}_1},\mathscr{P}_2) \rightarrow \mathscr{P}_2$ & +25\% & \multirow{2}{*}{1} & \multirow{2}{*}{$\mathscr{P}_1 \rightarrow (\mathscr{P}_1) \rightarrow \mathscr{P}_2$} & \multirow{2}{*}{-33\%} \\ \cline{3-6}

& & $\mathscr{P}_3 - \mathscr{P}_{16}$ & 4 & $\mathscr{P}_1 \rightarrow (\mathscr{P}_1,\mathscr{P}_1,\pmb{\mathscr{P}_3},\pmb{\mathscr{P}_4}) \rightarrow \mathscr{P}_2$ & +116\% & & \\ \hline
\multirow{2}{*}{$\mathscr{P}_1 \rightarrow (\mathscr{P}_3,\mathscr{P}_5,\mathscr{P}_6) \rightarrow \mathscr{P}_2$} & \multirow{2}{*}{0.77} & $\mathscr{P}_2-\mathscr{P}_{16}$ & 4 &  $\mathscr{P}_1 \rightarrow (\pmb{\mathscr{P}_2},\mathscr{P}_3,\mathscr{P}_5,\mathscr{P}_6) \rightarrow \mathscr{P}_{2}$ & +13\% & 2 & $\mathscr{P}_1 \rightarrow (\mathscr{P}_3,\mathscr{P}_5) \rightarrow \mathscr{P}_2$ & -9\% \\ \cline{3-9}

~& ~ & $\mathscr{P}_3-\mathscr{P}_{16}$ & 5 & $\mathscr{P}_1 \rightarrow (\mathscr{P}_3,\pmb{\mathscr{P}_3},\pmb{\mathscr{P}_4},\mathscr{P}_5,\mathscr{P}_6) \rightarrow \mathscr{P}_2$ & +30\% & 1 & $\mathscr{P}_1 \rightarrow (\mathscr{P}_3) \rightarrow \mathscr{P}_2$ & -36\%  \\ \hline

\end{tabular}
\end{table*}

\section{Conclusion}

Today, the integration of NFV and SDN introduces a new distinguished, yet challenging networking architecture. One of these challenges is VNF over and underloads in the network during their operational time. This problem could increase resource wastage, SLA violations, and also decrease efficiency. In this paper, we conducted a theoretical study of the VNF scaling problem with the aim of minimizing deployment and forwarding costs in a datacenter. In fact, we addressed the aforementioned issue by presenting a jointly traffic load balancing and VNF scaling mechanism to overcome both over and underload phenomena in datacenters. In this regard, we formulated the problem as a MILP optimization model. Also, to cope with the time complexity of MILP, we relaxed it into two LP models. However, in LDCs, there are huge number of network switches and servers, that increase the number of constraints and variables in LP models significantly. Hence, the proposed central LP model is practically useless for LDCs. Therefore, we extend our approach to a distributed method based on the ADMM technique. However, the conventional ADMM is guaranteed to converge for two-block variables problem, while for multi-block variable problems, updating different variables sequentially has no convergence guarantee. Hence, because our problem is formulated in a multi-block variable form, we used an extension of the ADMM technique, called Randomly Permuted ADMM (RP-ADMM) to solve the problem in a distributed manner. To the best of our knowledge, our work is the first one to tackle this problem using a distributed optimization approach. We evaluated the performance of the presented models numerically in different scenarios based on k-fat-tree topology. The results validated the performance of our proposed algorithms. The most important part of our future work is to implement the introduced distributed approach in real-world test beds. On the other hand, considering a workload prediction module in our architecture is worth to be explored in future. In this way, we would be able to predict the traffic load and adjust VNF placement in advance. Additionally, the performance (e.g., end-to-end delay) is also a critical factor in forwarding path selection of service chains. In fact, it would influence the selection of candidate PMs and VMs. Therefore, it could be considered as an extension to the mathematical formulation as another future direction.

\section{Acknowledgement}
We are grateful to Islamic Azad University, Mashhad branch authorities, for their useful collaboration. This work has been partly supported by the German Federal Ministry of Education and Research (BMBF) under the project SEcure Networking for a DATa center cloud in Europe (SENDATE) (Project ID 16KIS0261).

\bibliographystyle{unsrt}
\bibliography{ref}

\end{document}